\documentclass[manuscript]{acmart}
\usepackage[nolist]{acronym}
\usepackage{hyperref}
\usepackage{subfig}
\usepackage{tabularx}
\usepackage{xcolor}

\AtBeginDocument{%
\providecommand\BibTeX{{%
\normalfont B\kern-0.5em{\scshape i\kern-0.25em b}\kern-0.8em\TeX}}}

\setcopyright{acmlicensed}
\begin{document}

\copyrightyear{2021} 
\acmYear{2021} 
\acmConference[AutomotiveUI '21 Adjunct]{13th International Conference on Automotive User Interfaces and Interactive Vehicular Applications}{September 9--14, 2021}{Leeds, United Kingdom}
\acmBooktitle{13th International Conference on Automotive User Interfaces and Interactive Vehicular Applications (AutomotiveUI '21 Adjunct), September 9--14, 2021, Leeds, United Kingdom}\acmDOI{10.1145/3473682.3480252}
\acmISBN{978-1-4503-8641-8/21/09}

\begin{acronym}[]
	\acro{ECU}{Electronic Control Unit}
	\acro{GDPR}{General Data Protection Regulation}
	\acro{HCI}{Human-Computer Interaction}
	\acro{HMI}{Human-Machine Interface}
	\acro{HU}{Head Unit}
	\acro{IVIS}{In-Vehicle Information System}
	\acro{KPI}{Key Performance Indicator}
	\acro{NHTSA}{National Highway Traffic Safety Administration}
	\acro{OEM}{Original Equipment Manufacturer}
	\acro{OTA}{Over-The-Air}
	\acro{RH}{Research Hub}
	\acro{SE}{Steering Entropy}
	\acro{SWRR}{Steering Wheel Reversal Rate}
	\acro{UCD}{User-centered Design}
	\acro{UX}{User Experience}
\end{acronym}

%
% The "title" command has an optional parameter, allowing the author to define a "short title" to be used in page headers.
\title[Measuring Interaction-based Secondary Task Load: A Large-Scale Approach using Real-World Driving Data]{Measuring Interaction-based Secondary Task Load: A Large-Scale Approach using Real-World Driving Data}

%
% The "author" command and its associated commands are used to define the authors and their affiliations.
% Of note is the shared affiliation of the first two authors, and the "authornote" and "authornotemark" commands
% used to denote shared contribution to the research.
\author{Patrick Ebel}
\email{ebel@cs.uni-koeln.de}
\orcid{0000-0002-4437-2821}
\affiliation{%
	\institution{University of Cologne}
	\city{Cologne}
	\state{Germany}}
	
\author{Christoph Lingenfelder}
\email{christoph.lingenfelder@daimler.com}
\affiliation{%
	\institution{MBition}
	\city{Berlin}
	\state{Germany}}

\author{Andreas Vogelsang}
\email{vogelsang@cs.uni-koeln.de}
\orcid{0000-0003-1041-0815}
\affiliation{%
	\institution{University of Cologne}
	\city{Cologne}
	\state{Germany}}

\renewcommand{\shortauthors}{Ebel et al.}
%
% By default, the full list of authors will be used in the page headers. Often, this list is too long, and will overlap
% other information printed in the page headers. This command allows the author to define a more concise list
% of authors' names for this purpose.
%\renewcommand{\shortauthors}{Ebel, Brokhausen, Vogelsang}

% The abstract is a short summary of the work to be presented in the article.
\begin{abstract}

Center touchscreens are the main \ac{HMI} between the driver and the vehicle. They are becoming, larger, increasingly complex and replace functions that could previously be controlled using haptic interfaces. To ensure that touchscreen \acp{HMI} can be operated safely, they are subject to strict regulations and elaborate test protocols. Those methods and user trials require fully functional prototypes and are expensive and time-consuming. Therefore it is desirable to estimate the workload of specific interfaces or interaction sequences as early as possible in the development process. To address this problem, we envision a model-based approach that, based on the combination of user interactions and UI elements, can predict the secondary task load of the driver when interacting with the center screen. In this work, we present our current status, preliminary results, and our vision for a model-based system build upon large-scale natural driving data.
\end{abstract}


% %	
% %
% % Keywords. The author(s) should pick words that accurately describe the work being
% % presented. Separate the keywords with commas.

%
% This command processes the author and affiliation and title information and builds
% the first part of the formatted document.
\maketitle

\section{Introduction}
According to the \ac{NHTSA}~\cite{NHTSA.2021} 3,142 people were killed and 424,000 people were injured in crashes where drivers were distracted from the main driving task. Although the number of driver monitoring systems has increased steadily in recent years, a decrease in the number of accidents due to distracted driving can't be observed. Whereas the usage of smartphones during driving plays a major role, there are concerns that the increase in complexity, capability, and size of touchscreen-based \acp{HMI} will additionally increase the driver's cognitive workload. To prevent automotive \acp{HMI} from being too distracting, they undergo expensive and time-consuming empirical testing before they can be integrated into production line vehicles. While such measures are essential and will remain necessary, early feedback on potentially distracting usage patterns can be valuable for UX experts to design systems that are safe to use. A system that can predict the secondary task load based on the anticipated UI interactions and their properties, such as the sequence in which they occur or position on the display, can help UX experts to detect potentially distracting designs at an early stage and to develop appropriate alternatives. To enable such predictions, we envision a system that is based on driving and interaction data, automatically collected from a large amount of production line vehicles. Having access to such large-scale data makes it possible to generate insights that go beyond the detail of current, mostly qualitative or relatively small-scale naturalistic driving studies \cite{Ebel.2020}. Additionally, as soon as a software update is deployed to the fleet, the changes that were made can directly be assessed. In this work, we first investigate how the engagement of drivers with the touch-based \ac{HMI} can be measured using driving parameters and UI interaction data. We further elaborate on specific UI features and behavior that might lead to increased driver distraction.

\section{Related Work}
Driver Distraction Measurement is a well-studied field of research that will remain relevant even in the approaching age of automated driving. In certain driving environments, drivers will need to take over control of the car for a long time. Therefore, multiple approaches exist that try to predict and model driver distraction. A large group of such approaches is based on physiological data \cite{Solovey.2014, Schneegass.2013, Gjoreski.2020, Aghaei.2016} or data retrieved from eye-tracking systems \cite{Palinko.2010, Engstroem.2005, Pettitt.2010}. Whereas many of these approaches provide promising results and have proven their capability to effectively detect distracted driving, multiple factors prevent their large-scale usage. The main drawback of approaches based on physiological data is the fact that sensors need either to be attached to the body of the driver, or additional measurement units need to be installed in the car. This makes it nearly impossible to apply such methods outside of experimental environments or naturalistic driving studies. For approaches based on eye tracking, the costs of highly accurate eye-tracking systems are still a limiting factor for widespread deployment in production vehicles. Due to the highly sensitive nature of the data, most \acp{OEM} are reluctant to store video or gaze data. In contrast, methods that are based on driving data \cite{Kircher.2010, Li.2018, Risteska.2018, Nakayama.1999, Markkula.2006, Salis.2019} such as steering wheel angle, speed deviations, or vehicle accelerations, are more suitable for large-scale use cases since the data is already available in all modern cars and no additional instrumentation is necessary. Already in 1999 \citeauthor{Nakayama.1999}~\cite{Nakayama.1999} introduced the so-called Steering Entropy metric and they were able to show clear correlations between an increase in driver workload and steering behavior. Additionally, \citeauthor{Markkula.2006}~\cite{Markkula.2006} introduced the \ac{SWRR} and compared it to other steering angle metrics concerning their sensitivity to the effect of secondary task workload on lateral control performance. \citeauthor{Markkula.2006} found that the \ac{SWRR} and the Steering Entropy are superior compared to other metrics. Whereas the original experiments and most of the studies that build upon the findings of \citeauthor{Nakayama.1999} and \citeauthor{Markkula.2006} are based on data retrieved from simulator studies, a recent approach by \citeauthor{Li.2018} shows a clear correlation between driver distraction and high steering entropy. However, even though the data was collected in a natural driving environment, only 16 drivers over 12 days contributed to the data collection and additional instrumentation was installed in the cars.

\section{Approach}
To model the effect, specific user interactions or usage patterns on the touchscreen \ac{HMI} have on secondary task load, a large amount of data is needed. This is due to the many UI elements and the even larger number of potential combinations in modern \acp{HMI} as well as due to the diverse range of driving situations for which the effect might be different. Collecting this large amount of data in the form of naturalistic driving studies is time-consuming and expensive. Therefore, in this work, we utilize data collected from production line vehicles. However, this results in limitations, as in contrast to laboratory studies, strict data protection regulations must be met. For this reason, it is not possible to collect personalized data. Additionally, due to the many different participants and the uncontrolled driving environment, the data quality is significantly different compared to simulator studies or controlled naturalistic driving studies. This makes a detailed data analysis and preprocessing necessary. In this work in progress, we present our current state of research containing the data collection and processing methods, first preliminary results, and our research agenda.

\subsection{Data Collection and Processing}
The data used in this work is collected via a telematic framework that allows live \ac{OTA} data transfer from the car to the backend where the data is processed. The framework is available in the new generation of production vehicles and no additional instrumentation is needed. Detailed descriptions of the telematics architecture, processing framework, and data collection are provided by \citeauthor{Ebel.2021}~\cite{Ebel.2021}.

\begin{table*}
	\caption{Processed Signals}
	\centering
	\label{tab:Datapoints}
	\small
	\begin{tabular}{@{}ccll@{}}
		\toprule
		Variable & Unit & Description & Frequency\\
		\midrule
		$v$ 	    & km\textbackslash h & Vehicle Speed 		                            & 5\,Hz\\
		$\theta$ 	& $^{\circ}$ 	    & Steering Wheel Angle                              & 5\,Hz\\
		S 	        & - 	            & ADAS Status (Cruise Control and Steering Assist)  & On Change\\
		UI 	        & - 	            & Touchscreen Event (UI Element and Gesture)        & On Change\\
	\end{tabular}
\end{table*}

In the first processing step, the interaction sequences are extracted. The event sequence data consists of timestamped events containing the name of the interactive UI element that was triggered by the user and the type of gesture that was detected. We consider an interaction sequence to be a sequence of interactions where the time interval between two consecutive interactions is less than $t_{max}=10\,s$. In the second step, the interaction data is enriched with the driving data shown in Table \ref{tab:Datapoints}. We only consider sequences in which the driver assistance systems were not active. For the remaining sequences, we also consider the driving data immediately preceding the first and immediately following the last interaction of a sequence.

This is based on the assumption that the anticipated interaction with the HMI, before the actual gesture on the touchscreen is made, already influences the drivers' driving behavior. The same applies to the driving behavior shortly after the last interaction, as drivers tend to wait for visual confirmation and then reach back to the steering wheel. Considering the findings made by \citeauthor{Large.2017}~\cite{Large.2017}, \citeauthor{Green.2015}~\cite{Green.2015} and \citeauthor{Pettitt.2010}~\cite{Pettitt.2010}, we choose a duration of $t_{\text{buffer}}=2\,s$. Applying the introduced prepossessing steps, 29,055 interaction sequences are extracted. To compare the driving behavior during \textit{interaction sequences} with the driving behavior during \textit{no-interaction sequences} we sampled the same amount of driving data snippets from sequences where no interactions were made. However, compared to a controlled experiment we define the no-interaction sequences such that during those periods no interactions were made with the in-vehicle HMI. We can't control for distractions happening outside of the head unit, for example, due to phone usage or passengers. To make a valid comparison between interaction sequences and no-interaction sequences we apply stratified sampling, such that the distribution in sequence length is equal in both groups.

After sequence extraction, aggregated statistics for each sequence as well as driver distraction metrics, namely \ac{SWRR} (1, 2, and 5 degree according to \citeauthor{Markkula.2006}~\cite{Markkula.2006}) and \ac{SE} (according to \citeauthor{Nakayama.1999}~\cite{Nakayama.1999} with adjustments proposed by \citeauthor{Boer.2005}~\cite{Boer.2005} to avoid extremely high entropies based on outliers) are calculated. Since no personalized data is available it is not possible to calculate a personalized $\alpha$ for the \ac{SE} metric. We, therefore, averaged over all no-interaction sequences to obtain an average $\alpha$.

\subsection{Preliminary Results}
Similar to \citeauthor{Markkula.2006}~\cite{Markkula.2006}, we compare the steering wheel metrics based on their standardized effect size $d$ \cite{Cohen.1988} and two different driving conditions. We differentiate between straight driving and curved driving since previous work \cite{Boer.2005, Markkula.2006, Nakayama.1999} found that the SE and SWRR metrics are highly sensitive with regard to the road curvature. 

	\begin{figure*}
		\centering
		\includegraphics[width=0.5\linewidth]{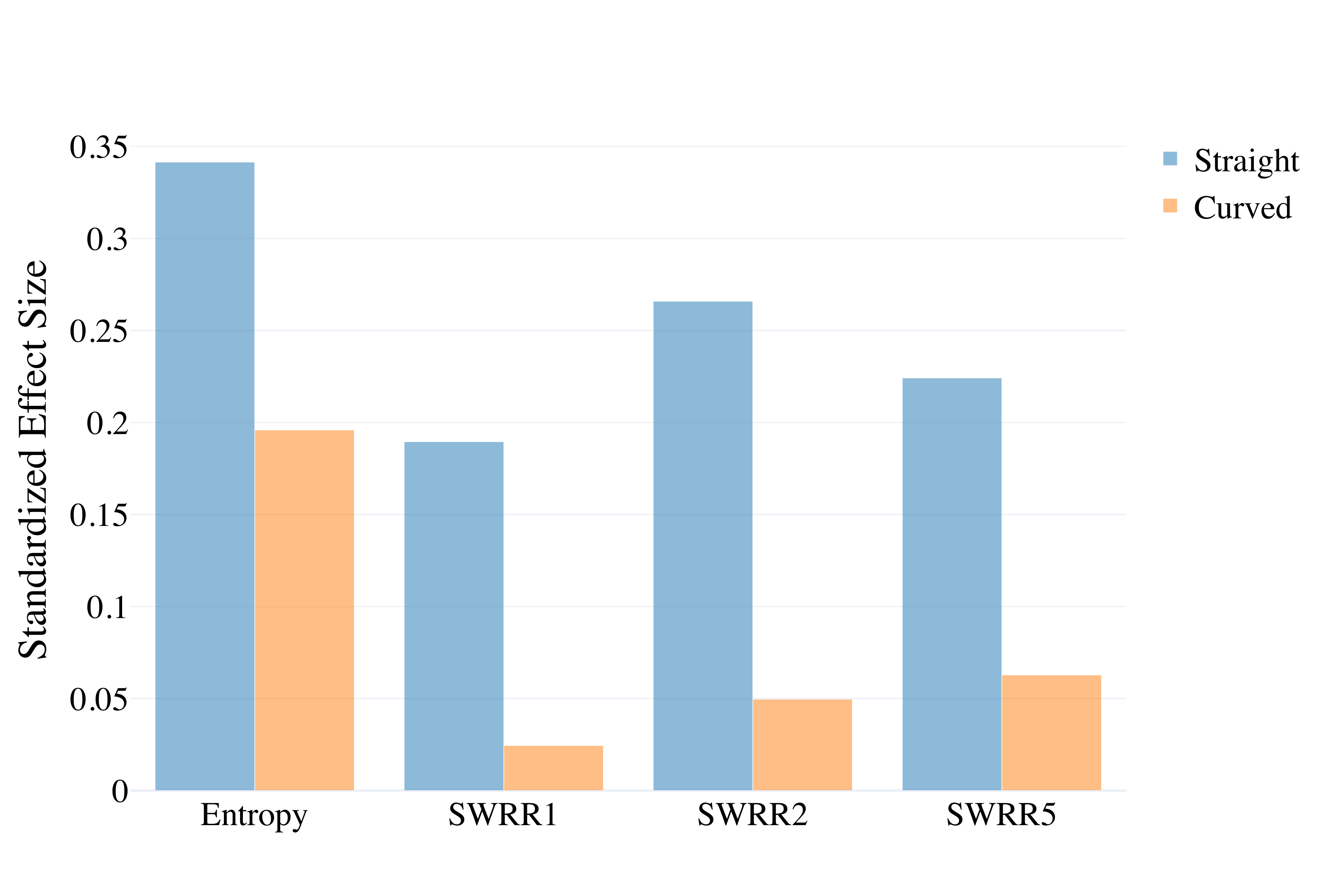} %Here goes the path to the figure (Best to use vector graphics (.pdf/.svg)
		\caption{Standardized effect sizes for the steering wheel metrics and driving conditions (straight, curved)} %Caption 
		\label{fig:effect} %Label to refer to this image in the text
	\end{figure*}
	
As shown in Fig. \ref{fig:effect} one can observe the difference in effect size between straight and curved driving, meaning that interaction and no-interaction sequences can be better separated for straight driving. The SWRR metrics are even more affected than the steering entropy. In general, the standardized effect sizes are smaller than reported by \citeauthor{Markkula.2006}~\cite{Markkula.2006} (e.g. $d_{SE} = 0.34$ for the data in this work compared to $d_{SE} = 0.8$). Multiple reasons may cause this difference. The first and probably most important difference is the driving context in which the data was collected. Whereas, the dataset at hand comprises of multiple different driving scenarios, drivers, and interactions, the data used by \citeauthor{Markkula.2006} was collected in a field study, where 48 participants drove the same sequence on a motorway performing the same two tasks (half of the participants performed a visual arrow task the other half a cognitive one). Additionally, we did not yet tune the metrics to increase sensitivity.

	\begin{figure*}

		\centering
		\subfloat[Steering Entropy]{\includegraphics[width=0.5\linewidth]{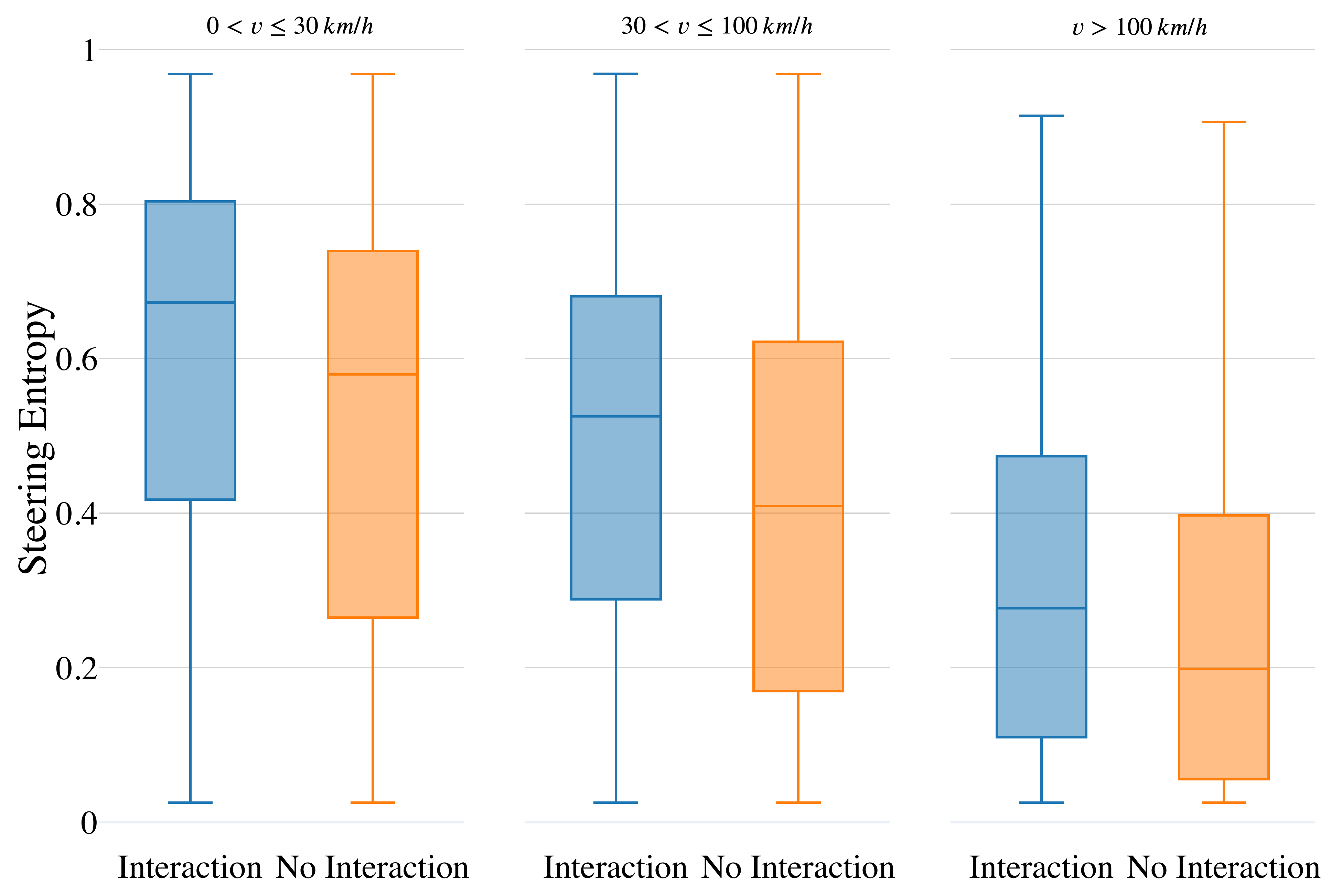}}
% 		 \qquad
        \subfloat[2$^{\circ}$ Steering Wheel Reversal Rate]{\includegraphics[width=0.5\linewidth]{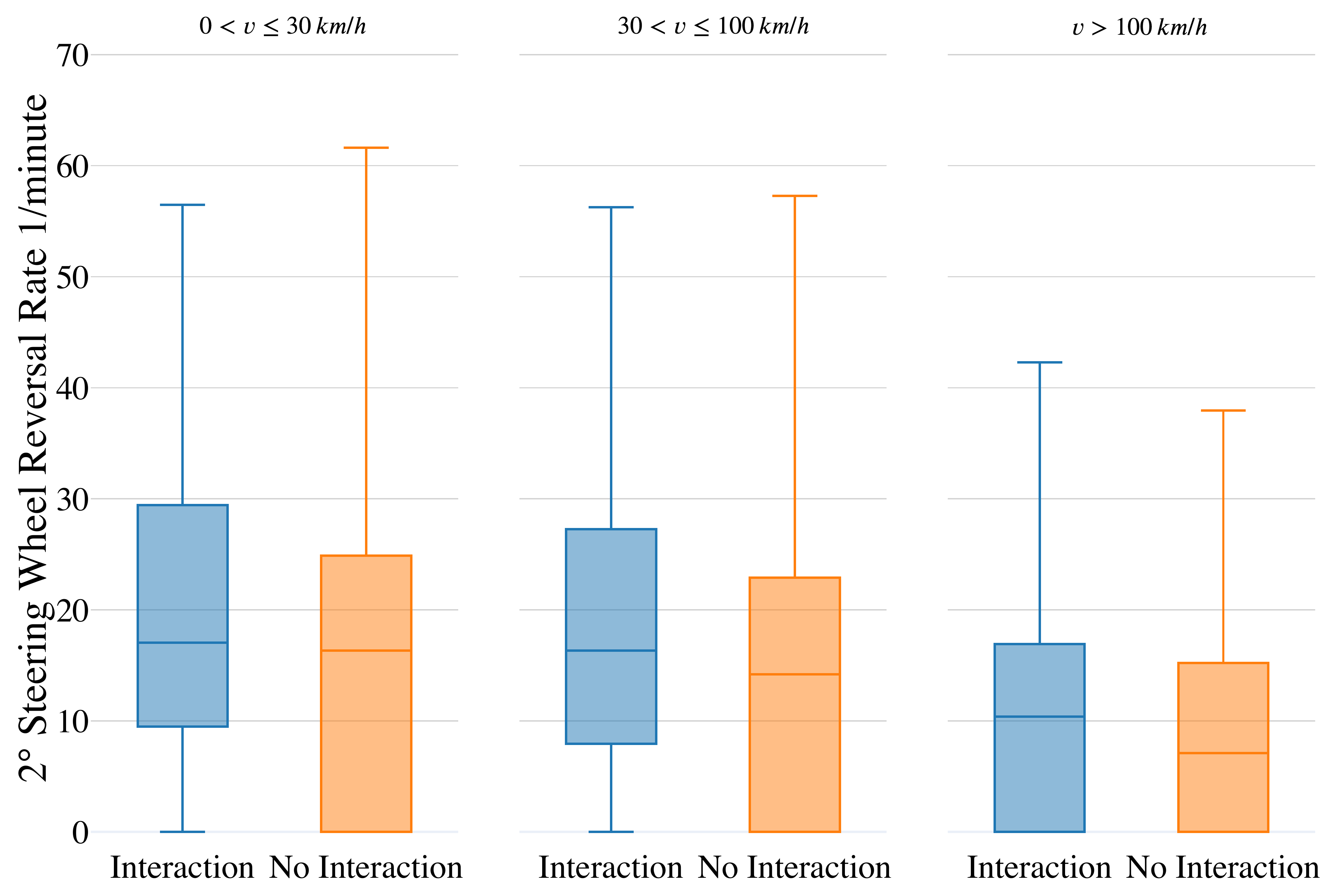}}
		\caption{Steering Entropy and Steering Wheel Reversal Rate for straight driving sequences with HMI Interaction and without HMI interaction for different speed buckets.}
		\label{fig:Comparison}
	\end{figure*}

In Fig. \ref{fig:Comparison} a comparison between interaction and no-interaction sequences for straight driving is presented. We report the SE and the 2$^{\circ}$ SWRR since they have been found to be most sensitive. Even though the driving environment is highly uncontrolled and multiple confounding factors can influence the driving behavior, one can clearly observe the anticipated differences between interaction and no-interaction sequences for the \ac{SWRR} and for the \ac{SE}. One can see that the \ac{SE} is larger for smaller vehicle speeds. Whereas the 2$^{\circ}$ and 5$^{\circ}$ \ac{SWRR} measures show the same effect, this trend is not observable in the 1$^{\circ}$ \ac{SWRR} (not displayed). A comparison of the 1$^{\circ}$ SWRR with the results from the field experiment conducted by \citeauthor{Engstroem.2005}~\cite{Engstroem.2005} lead to similar results in the absolute values and the difference between interaction sequences and no-interaction sequences.

\subsection{Research Agenda}
The preliminary results show that, even in the highly uncontrolled setting of real-world data, both metrics studied are suitable to measure secondary task load induced by HU touch interactions. In particular, the SE provided promising results in terms of sensitivity. As a first next step, we plan to adjust the metrics to increase sensitivity. Then, building on the current state of work, we plan to evaluate the correlation between certain UI elements, interactions or interaction patterns, and the driver distraction metrics. Proving this correlation is an important step toward a predictive model of secondary task load. To then draw conclusions about the workload induced by specific interactions with certain UI elements, we plan to perform a feature importance analysis. First, we will use a machine learning-based approach to predict steering entropy based on the proportion of different interactions (e.g. list tab or map drag) and additional metadata like the number of interactions and interaction density. The insights generated via a feature importance analysis can then serve as a first feedback on interactions and interaction patterns that highly influence secondary task load. The overarching goal is to develop an evaluation tool, that 

\section{Discussion}
In this work in progress we present our planned approach to develop a model-based method, leveraging real-world data to predict secondary task load induced by interactions with a touchscreen \ac{HMI}. The predictions should be based on the type of interaction and the respective type of UI element used and can serve as an early-stage estimate of driver distraction far before the first experiments are conducted. Therefore, UX experts get an early feedback on their designs which supports them to design non-distracting interfaces that increase road safety and to save costs since re-designs due to necessary changes in later studies can be avoided.

\bibliographystyle{ACM-Reference-Format}
\bibliography{AutoUI21WIP.bib}

\end{document}